\def\temp{1.35}%
\let\tempp=\relax
\expandafter\ifx\csname psboxversion\endcsname\relax
\else
    \ifdim\temp cm>\psboxversion cm
    \else
      \let\temp=\psboxversion
      \let\tempp= 
    \fi
\fi
\tempp
\let\psboxversion=\temp
\catcode`\@=11
\def\psfortextures{%     For TeXtures on the Macintosh
\def\PSspeci@l##1##2{%
\special{illustration ##1\space scaled ##2}%
}}%
\def\psfordvitops{%      For the DVItoPS converter on IBM mainframes
\def\PSspeci@l##1##2{%
\special{dvitops: import ##1\space \the\drawingwd \the\drawinght}%
}}%
\def\psfordvips{%      For DVIPS converter on VAX, UNIX and PC's
\def\PSspeci@l##1##2{%
\d@my=0.1bp \d@mx=\drawingwd \divide\d@mx by\d@my% BUG! for large \drawingwd
\includegraphics{##1\space}}}%
\def\psforoztex{%        For the OzTeX shareware on the Macintosh
\def\PSspeci@l##1##2{%
\special{##1 \space
      ##2 1000 div dup scale
      \number-\psllx\space\space \number-\pslly\space\space translate
}}}%
\def\psfordvitps{%       From the UNIX TeXPS package, vers.>3.12
\def\dvitpsLiter@ldim##1{\dimen0=##1\relax
\special{dvitps: Literal "\number\dimen0\space"}}%
\def\PSspeci@l##1##2{%
\at(0bp;\drawinght){%
\special{dvitps: Include0 "psfig.psr"}% contains def of "startTexFig"
\dvitpsLiter@ldim{\drawingwd}%
\dvitpsLiter@ldim{\drawinght}%
\dvitpsLiter@ldim{\psllx bp}%
\dvitpsLiter@ldim{\pslly bp}%
\dvitpsLiter@ldim{\psurx bp}%
\dvitpsLiter@ldim{\psury bp}%
\special{dvitps: Literal "startTexFig"}%
\special{dvitps: Include1 "##1"}%
\special{dvitps: Literal "endTexFig"}%
}}}%
\def\psfordvialw{%   Try for dvialw, a UNIX public domain
\def\PSspeci@l##1##2{
\special{language "PostScript",
position = "bottom left",
literal "  \psllx\space \pslly\space translate
  ##2 1000 div dup scale
  -\psllx\space -\pslly\space translate",
include "##1"}
}}%
\def\psforptips{%   For MS-DOS; LUOMA@brandeis.bitnet
\def\PSspeci@l##1##2{{
\d@mx=\psurx bp
\advance \d@mx by -\psllx bp
\divide \d@mx by 1000\multiply\d@mx by \xscale
\incm{\d@mx}
\let\tmpx\dimincm
\d@my=\psury bp
\advance \d@my by -\pslly bp
\divide \d@my by 1000\multiply\d@my by \xscale
\incm{\d@my}
\let\tmpy\dimincm
\d@mx=-\psllx bp
\divide \d@mx by 1000\multiply\d@mx by \xscale
\d@my=-\pslly bp
\divide \d@my by 1000\multiply\d@my by \xscale
\at(\d@mx;\d@my){\special{ps:##1 x=\tmpx cm, y=\tmpy cm}}
}}}%
\def\psonlyboxes{%     Draft-like behaviour if none of the others works
\def\PSspeci@l##1##2{%
\at(0cm;0cm){\boxit{\vbox to\drawinght
  {\vss\hbox to\drawingwd{\at(0cm;0cm){\hbox{({\tt##1})}}\hss}}}}
}}%
\def\psloc@lerr#1{%
\let\savedPSspeci@l=\PSspeci@l%
\def\PSspeci@l##1##2{%
\at(0cm;0cm){\boxit{\vbox to\drawinght
  {\vss\hbox to\drawingwd{\at(0cm;0cm){\hbox{({\tt##1}) #1}}\hss}}}}
\let\PSspeci@l=\savedPSspeci@l% restore normal output for other figs!
}}%
\newread\pst@mpin
\newdimen\drawinght\newdimen\drawingwd
\newdimen\psxoffset\newdimen\psyoffset
\newbox\drawingBox
\newcount\xscale \newcount\yscale \newdimen\pscm\pscm=1cm
\newdimen\d@mx \newdimen\d@my
\newdimen\pswdincr \newdimen\pshtincr
\let\ps@nnotation=\relax
{\catcode`\|=0 |catcode`|\=12 |catcode`|%=12 |catcode`~=12
|catcode`#=12 |catcode`*=14
|xdef|backslashother{\}*
|xdef|percentother{%}*
|xdef|tildeother{~}*
|xdef|sharpother{#}*
}%
\def\R@moveMeaningHeader#1:->{}%
\def\uncatcode#1{%
\edef#1{\expandafter\R@moveMeaningHeader\meaning#1}}%
\def\execute#1{#1}% NOT stupid: cs in #1 are then identified BEFORE execution
\def\psm@keother#1{\catcode`#112\relax}% borrowed from latex
\def\executeinspecs#1{%
\execute{\begingroup\let\do\psm@keother\dospecials\catcode`\^^M=9#1\endgroup}}%
\def\@mpty{}%
\def\matchexpin#1#2{
  \fi%
  \edef\tmpb{{#2}}%
  \expandafter\makem@tchtmp\tmpb%
  \edef\tmpa{#1}\edef\tmpb{#2}%
  \expandafter\expandafter\expandafter\m@tchtmp\expandafter\tmpa\tmpb\endm@tch%
  \if\match%
}%
\def\matchin#1#2{%
  \fi%
  \makem@tchtmp{#2}%
  \m@tchtmp#1#2\endm@tch%
  \if\match%
}%
\def\makem@tchtmp#1{\def\m@tchtmp##1#1##2\endm@tch{%
  \def\tmpa{##1}\def\tmpb{##2}\let\m@tchtmp=\relax%
  \ifx\tmpb\@mpty\def\match{YN}%
  \else\def\match{YY}\fi%
}}%
\def\incm#1{{\psxoffset=1cm\d@my=#1
 \d@mx=\d@my
  \divide\d@mx by \psxoffset
  \xdef\dimincm{\number\d@mx.}
  \advance\d@my by -\number\d@mx cm
  \multiply\d@my by 100
 \d@mx=\d@my
  \divide\d@mx by \psxoffset
  \edef\dimincm{\dimincm\number\d@mx}
  \advance\d@my by -\number\d@mx cm
  \multiply\d@my by 100
 \d@mx=\d@my
  \divide\d@mx by \psxoffset
  \xdef\dimincm{\dimincm\number\d@mx}
}}%
\newif\ifNotB@undingBox
\newhelp\PShelp{Proceed: you'll have a 5cm square blank box instead of
your graphics.}%
\def\s@tsize#1 #2 #3 #4\@ndsize{
  \def\psllx{#1}\def\pslly{#2}%
  \def\psurx{#3}\def\psury{#4}%  needed by a crazyness of dvips!
  \ifx\psurx\@mpty\NotB@undingBoxtrue% this is not a valid one!
  \else
    \drawinght=#4bp\advance\drawinght by-#2bp
    \drawingwd=#3bp\advance\drawingwd by-#1bp
  \fi
  }%
\def\sc@nBBline#1:#2\@ndBBline{\edef\p@rameter{#1}\edef\v@lue{#2}}%
\def\g@bblefirstblank#1#2:{\ifx#1 \else#1\fi#2}%
{\catcode`\%=12
\xdef\B@undingBox{%%BoundingBox}}%
\def\ReadPSize#1{
 \readfilename#1\relax
 \let\PSfilename=\lastreadfilename
 \openin\pst@mpin=#1\relax
 \ifeof\pst@mpin \errhelp=\PShelp
   \errmessage{I haven't found your postscript file (\PSfilename)}%
   \psloc@lerr{was not found}%
   \s@tsize 0 0 142 142\@ndsize
   \closein\pst@mpin
 \else
   \if\matchexpin{\GlobalInputList}{, \lastreadfilename}%
   \else\xdef\GlobalInputList{\GlobalInputList, \lastreadfilename}%
     \immediate\write\psbj@inaux{\lastreadfilename,}%
   \fi%
   \loop
     \executeinspecs{\catcode`\ =10\global\read\pst@mpin to\n@xtline}%
     \ifeof\pst@mpin
       \errhelp=\PShelp
       \errmessage{(\PSfilename) is not an Encapsulated PostScript File:
           I could not find any \B@undingBox: line.}%
       \edef\v@lue{0 0 142 142:}%
       \psloc@lerr{is not an EPSFile}%
       \NotB@undingBoxfalse
     \else
       \expandafter\sc@nBBline\n@xtline:\@ndBBline
       \ifx\p@rameter\B@undingBox\NotB@undingBoxfalse
         \edef\t@mp{%
           \expandafter\g@bblefirstblank\v@lue\space\space\space}%
         \expandafter\s@tsize\t@mp\@ndsize
       \else\NotB@undingBoxtrue
       \fi
     \fi
   \ifNotB@undingBox\repeat
   \closein\pst@mpin
 \fi
\message{#1}%
}%
\def\psboxto(#1;#2)#3{\vbox{%
   \ReadPSize{#3}%
   \advance\pswdincr by \drawingwd
   \advance\pshtincr by \drawinght
   \divide\pswdincr by 1000
   \divide\pshtincr by 1000
   \d@mx=#1
   \ifdim\d@mx=0pt\xscale=1000
         \else \xscale=\d@mx \divide \xscale by \pswdincr\fi
   \d@my=#2
   \ifdim\d@my=0pt\yscale=1000
         \else \yscale=\d@my \divide \yscale by \pshtincr\fi
   \ifnum\yscale=1000
         \else\ifnum\xscale=1000\xscale=\yscale
                    \else\ifnum\yscale<\xscale\xscale=\yscale\fi
              \fi
   \fi
   \divide\drawingwd by1000 \multiply\drawingwd by\xscale
   \divide\drawinght by1000 \multiply\drawinght by\xscale
   \divide\psxoffset by1000 \multiply\psxoffset by\xscale
   \divide\psyoffset by1000 \multiply\psyoffset by\xscale
   \global\divide\pscm by 1000
   \global\multiply\pscm by\xscale
   \multiply\pswdincr by\xscale \multiply\pshtincr by\xscale
   \ifdim\d@mx=0pt\d@mx=\pswdincr\fi
   \ifdim\d@my=0pt\d@my=\pshtincr\fi
   \message{scaled \the\xscale}%
 \hbox to\d@mx{\hss\vbox to\d@my{\vss
   \global\setbox\drawingBox=\hbox to 0pt{\kern\psxoffset\vbox to 0pt{%
      \kern-\psyoffset
      \PSspeci@l{\PSfilename}{\the\xscale}%
      \vss}\hss\ps@nnotation}%
   \global\wd\drawingBox=\the\pswdincr
   \global\ht\drawingBox=\the\pshtincr
   \global\drawingwd=\pswdincr
   \global\drawinght=\pshtincr
   \baselineskip=0pt
   \copy\drawingBox
 \vss}\hss}%
  \global\psxoffset=0pt
  \global\psyoffset=0pt
  \global\pswdincr=0pt
  \global\pshtincr=0pt % These are local to one figure
  \global\pscm=1cm %should not be necessary
}}%
\def\psboxscaled#1#2{\vbox{%
  \ReadPSize{#2}%
  \xscale=#1
  \message{scaled \the\xscale}%
  \divide\pswdincr by 1000 \multiply\pswdincr by \xscale
  \divide\pshtincr by 1000 \multiply\pshtincr by \xscale
  \divide\psxoffset by1000 \multiply\psxoffset by\xscale
  \divide\psyoffset by1000 \multiply\psyoffset by\xscale
  \divide\drawingwd by1000 \multiply\drawingwd by\xscale
  \divide\drawinght by1000 \multiply\drawinght by\xscale
  \global\divide\pscm by 1000
  \global\multiply\pscm by\xscale
  \global\setbox\drawingBox=\hbox to 0pt{\kern\psxoffset\vbox to 0pt{%
     \kern-\psyoffset
     \PSspeci@l{\PSfilename}{\the\xscale}%
     \vss}\hss\ps@nnotation}%
  \advance\pswdincr by \drawingwd
  \advance\pshtincr by \drawinght
  \global\wd\drawingBox=\the\pswdincr
  \global\ht\drawingBox=\the\pshtincr
  \global\drawingwd=\pswdincr
  \global\drawinght=\pshtincr
  \baselineskip=0pt
  \copy\drawingBox
  \global\psxoffset=0pt
  \global\psyoffset=0pt
  \global\pswdincr=0pt
  \global\pshtincr=0pt % These are local to one figure
  \global\pscm=1cm
}}%
\def\psbox#1{\psboxscaled{1000}{#1}}%
\newif\ifn@teof\n@teoftrue
\newif\ifc@ntrolline
\newif\ifmatch
\newread\j@insplitin
\newwrite\j@insplitout
\newwrite\psbj@inaux
\immediate\openout\psbj@inaux=psbjoin.aux
\immediate\write\psbj@inaux{\string\joinfiles}%
\immediate\write\psbj@inaux{\jobname,}%
\def\toother#1{\ifcat\relax#1\else\expandafter%
  \toother@ux\meaning#1\endtoother@ux\fi}%
\def\toother@ux#1 #2#3\endtoother@ux{\def\tmp{#3}%
  \ifx\tmp\@mpty\def\tmp{#2}\let\next=\relax%
  \else\def\next{\toother@ux#2#3\endtoother@ux}\fi%
\next}%
\let\readfilenamehook=\relax
\def\re@d{\expandafter\re@daux}% spares typing 10 \expandafter's...
\def\re@daux{\futurelet\nextchar\stopre@dtest}%
\def\re@dnext{\xdef\lastreadfilename{\lastreadfilename\nextchar}%
  \afterassignment\re@d\let\nextchar}%
\def\stopre@d{\egroup\readfilenamehook}%
\def\stopre@dtest{%
  \ifcat\nextchar\relax\let\nextread\stopre@d
  \else
    \ifcat\nextchar\space\def\nextread{%
      \afterassignment\stopre@d\chardef\nextchar=`}%
    \else\let\nextread=\re@dnext
      \toother\nextchar
      \edef\nextchar{\tmp}%
    \fi
  \fi\nextread}%
\def\readfilename{\bgroup%
  \let\\=\backslashother \let\%=\percentother \let\~=\tildeother
  \let\#=\sharpother \xdef\lastreadfilename{}%
  \re@d}%
\xdef\GlobalInputList{\jobname}%
\def\psnewinput{%
  \def\readfilenamehook{% each entry in \GlobalInputList should be unique
    \if\matchexpin{\GlobalInputList}{, \lastreadfilename}%
    \else\xdef\GlobalInputList{\GlobalInputList, \lastreadfilename}%
      \immediate\write\psbj@inaux{\lastreadfilename,}%
    \fi%
    \let\readfilenamehook=\relax%
    \ps@ldinput\lastreadfilename\relax%
  }\readfilename%
}%
\expandafter\ifx\csname @@input\endcsname\relax    % then Plain
  \immediate\let\ps@ldinput=\input\def\input{\psnewinput}%
\else
  \immediate\let\ps@ldinput=\@@input
  \def\@@input{\psnewinput}%
\fi%
\def\nowarnopenout{%
 \def\warnopenout##1##2{%
   \readfilename##2\relax
   \message{\lastreadfilename}%
   \immediate\openout##1=\lastreadfilename\relax}}%
\def\warnopenout#1#2{%
 \readfilename#2\relax
 \def\t@mp{TrashMe,psbjoin.aux,psbjoint.tex,}\uncatcode\t@mp
 \if\matchexpin{\t@mp}{\lastreadfilename,}%
 \else
   \immediate\openin\pst@mpin=\lastreadfilename\relax
   \ifeof\pst@mpin
     \else
     \edef\tmp{{If the content of this file is precious to you, this
is your last chance to abort (ie press x or e) and rename it before
retexing (\jobname). If you're sure there's no file
(\lastreadfilename) in the directory of (\jobname), then go on: I'm
simply worried because you have another (\lastreadfilename) in some
directory I'm looking in for inputs...}}%
     \errhelp=\tmp
     \errmessage{I may be about to replace your file named \lastreadfilename}%
   \fi
   \immediate\closein\pst@mpin
 \fi
 \message{\lastreadfilename}%
 \immediate\openout#1=\lastreadfilename\relax}%
{\catcode`\%=12\catcode`\*=14
\gdef\splitfile#1{*
 \readfilename#1\relax
 \immediate\openin\j@insplitin=\lastreadfilename\relax
 \ifeof\j@insplitin
   \message{! I couldn't find and split \lastreadfilename!}*
 \else
   \immediate\openout\j@insplitout=TrashMe
   \message{< Splitting \lastreadfilename\space into}*
   \loop
     \ifeof\j@insplitin
       \immediate\closein\j@insplitin\n@teoffalse
     \else
       \n@teoftrue
       \executeinspecs{\global\read\j@insplitin to\spl@tinline\expandafter
         \ch@ckbeginnewfile\spl@tinline%Beginning-Of-File-Named:%\endcheck}*
       \ifc@ntrolline
       \else
         \toks0=\expandafter{\spl@tinline}*
         \immediate\write\j@insplitout{\the\toks0}*
       \fi
     \fi
   \ifn@teof\repeat
   \immediate\closeout\j@insplitout
 \fi\message{>}*
}*
\gdef\ch@ckbeginnewfile#1%Beginning-Of-File-Named:#2%#3\endcheck{*
 \def\t@mp{#1}*
 \ifx\@mpty\t@mp
   \def\t@mp{#3}*
   \ifx\@mpty\t@mp
     \global\c@ntrollinefalse
   \else
     \immediate\closeout\j@insplitout
     \warnopenout\j@insplitout{#2}*
     \global\c@ntrollinetrue
   \fi
 \else
   \global\c@ntrollinefalse
 \fi}*
\gdef\joinfiles#1\into#2{*
 \message{< Joining following files into}*
 \warnopenout\j@insplitout{#2}*
 \message{:}*
 {*
 \edef\w@##1{\immediate\write\j@insplitout{##1}}*
\w@{% This collection of files was produced with CERN psbox package}*
\w@{% To decompose and tex it:}*
\w@{%-save this with a filename CONTAINING ONLY LETTERS and a .TEX}*
\w@{% extension (say, JOINTFIL.TEX), in some empty directory;}*
\w@{%-make sure you can \string\input\space psbox.tex (version>=1.3);}*
\w@{%  (else ftp cs.nyu.edu(=128.122.140.24):pub/TeX/psbox/, then get}*
\w@{%  and tex the file psboxall.tex; more info in psbREAD.ME)}*
\w@{%-tex JOINTFIL.TEX using Plain, or LaTeX, or whatever is needed by}*
\w@{%  the first file in the joining (after splitting JOINTFIL.TEX into}*
\w@{%  it's constituents, TeX will try to process it as it stands).}*
\w@{\string\input\space psbox.tex}*
\w@{\string\splitfile{\string\jobname}}*
\w@{\string\let\string\autojoin=\string\relax}*
}*
 \expandafter\tre@tfilelist#1, \endtre@t
 \immediate\closeout\j@insplitout
 \message{>}*
}*
\gdef\tre@tfilelist#1, #2\endtre@t{*
 \readfilename#1\relax
 \ifx\@mpty\lastreadfilename
 \else
   \immediate\openin\j@insplitin=\lastreadfilename\relax
   \ifeof\j@insplitin
     \errmessage{I couldn't find file \lastreadfilename}*
   \else
     \message{\lastreadfilename}*
     \immediate\write\j@insplitout{%Beginning-Of-File-Named:\lastreadfilename}*
     \executeinspecs{\global\read\j@insplitin to\oldj@ininline}*
     \loop
       \ifeof\j@insplitin\immediate\closein\j@insplitin\n@teoffalse
       \else\n@teoftrue
         \executeinspecs{\global\read\j@insplitin to\j@ininline}*
         \toks0=\expandafter{\oldj@ininline}*
         \let\oldj@ininline=\j@ininline
         \immediate\write\j@insplitout{\the\toks0}*
       \fi
     \ifn@teof
     \repeat
   \immediate\closein\j@insplitin
   \fi
   \tre@tfilelist#2, \endtre@t
 \fi}*
}%
\def\autojoin{%
 \immediate\write\psbj@inaux{\string\into{psbjoint.tex}}%
 \immediate\closeout\psbj@inaux
 \expandafter\joinfiles\GlobalInputList\into{psbjoint.tex}%
}%
\def\centinsert#1{\midinsert\line{\hss#1\hss}\endinsert}%
\def\psannotate#1#2{\vbox{%
  \def\ps@nnotation{#2\global\let\ps@nnotation=\relax}#1}}%
\def\pscaption#1#2{\vbox{%
   \setbox\drawingBox=#1
   \copy\drawingBox
   \vskip\baselineskip
   \vbox{\hsize=\wd\drawingBox\setbox0=\hbox{#2}%
     \ifdim\wd0>\hsize
       \noindent\unhbox0\tolerance=5000
    \else\centerline{\box0}%
    \fi
}}}%
\def\at(#1;#2)#3{\setbox0=\hbox{#3}\ht0=0pt\dp0=0pt
  \rlap{\kern#1\vbox to0pt{\kern-#2\box0\vss}}}%
\newdimen\gridht \newdimen\gridwd
\def\gridfill(#1;#2){%
  \setbox0=\hbox to 1\pscm
  {\vrule height1\pscm width.4pt\leaders\hrule\hfill}%
  \gridht=#1
  \divide\gridht by \ht0
  \multiply\gridht by \ht0
  \gridwd=#2
  \divide\gridwd by \wd0
  \multiply\gridwd by \wd0
  \advance \gridwd by \wd0
  \vbox to \gridht{\leaders\hbox to\gridwd{\leaders\box0\hfill}\vfill}}%
\def\fillinggrid{\at(0cm;0cm){\vbox{%
  \gridfill(\drawinght;\drawingwd)}}}%
\def\textleftof#1:{%
  \setbox1=#1
  \setbox0=\vbox\bgroup
    \advance\hsize by -\wd1 \advance\hsize by -2em}%
\def\textrightof#1:{%
  \setbox0=#1
  \setbox1=\vbox\bgroup
    \advance\hsize by -\wd0 \advance\hsize by -2em}%
\def\endtext{%
  \egroup
  \hbox to \hsize{\valign{\vfil##\vfil\cr%
\box0\cr%
\noalign{\hss}\box1\cr}}}%
\def\frameit#1#2#3{\hbox{\vrule width#1\vbox{%
  \hrule height#1\vskip#2\hbox{\hskip#2\vbox{#3}\hskip#2}%
        \vskip#2\hrule height#1}\vrule width#1}}%
\def\boxit#1{\frameit{0.4pt}{0pt}{#1}}%
\catcode`\@=12 % cs containing @ are unreachable
\psfordvips   %     For DVIPS converter on VAX and UNIX
\documentstyle[]{mn}

\begin{document}

\title[Acceleration time scale in relativistic shock waves]{Acceleration
time scale for the first-order Fermi acceleration in relativistic shock
waves.}

\author[J. Bednarz \& M. Ostrowski]{J. Bednarz$^1$ \& M. Ostrowski$^{1,2}$ \\
$^1$Obserwatorium Astronomiczne, Uniwersytet Jagiello\'{n}ski, Krak\'{o}w,
Poland (E-mail: bednarz{@}oa.uj.edu.pl and  mio{@}oa.uj.edu.pl) \\
$^2$Max-Planck-Institut f\"ur Radiostronomie, Bonn, Germany}

\date{}

\maketitle

\begin{abstract}
The acceleration time scale for the process of first-order Fermi
acceleration in relativistic shock waves with oblique magnetic field
configurations is investigated by the method of Monte Carlo particle
simulations. We discuss the differences in derivation of the cosmic ray
acceleration time scale for non-relativistic and relativistic shocks. We
demonstrate the presence of correlation between the particle energy gain
at interaction with the shock and the respective time elapsed since the
previous interaction. Because of that any derivation of the acceleration
time scale can not use the distribution of energy gains and the
distribution of times separately. The time scale discussed in the
present paper, $T_{acc}^{(c)}$, is the one describing the rate of change
of the particle spectrum cut-off energy in the time dependent evolution.
It is derived using a simplified method involving small amplitude
particle momentum scattering and intended to model the situations with
anisotropic cosmic ray distributions. We consider shocks with parallel,
as well as oblique, sub- and super-luminal magnetic field configurations
with finite amplitude perturbations, $\delta B$.  At parallel shocks
$T_{acc}^{(c)}$ diminishes with the growing perturbation amplitude and
shock velocity $U_1$. Another feature discovered in oblique shocks are
non-monotonic changes of $T_{acc}^{(c)}$ with $\delta B$. The effect
arises due to the particle cross-field diffusion. The acceleration
process leading to power-law spectra is possible in super-luminal shocks
only in the presence of large amplitude turbulence. Then,
$T_{acc}^{(c)}$ always {\em increases} with increasing $\delta B$. In
some of the considered shocks the acceleration time scale can be shorter
than the particle gyroperiod upstream the shock. We also indicate the
relation existing for relativistic shocks between the acceleration time
scale and the particle spectral index. A short discussion of the
numerical approach modelling the pitch angle diffusion versus the large
angle momentum scattering is given. We stress the importance of the
proper evaluation of the effective magnetic field (including the
perturbed component) in simulations involving discrete particle momentum
scattering.
\end{abstract}


\section{Introduction}
 
A consistent method to tackle the problem of first-order Fermi
acceleration in relativistic shock waves was conceived by Kirk \&
Schneider (1987a;  see also Kirk 1988). By extending the diffusion
approximation to higher order terms in the anisotropy of particle
distribution, they obtained solutions to a kinetic  equation of the
Fokker--Planck type with the isotropic form of pitch  angle diffusion
coefficient. Next, Kirk \&  Schneider (1988) extended the analysis  by
involving  both diffusion  and large-angle scattering in particle  pitch
angle. They discovered that --  in relativistic shock waves -- the
presence of scattering  can substantially modify the spectrum of
accelerated particles. An extension of  Kirk \& Schneider's (1987a)
approach to more general conditions in the shock was given by Heavens \&
Drury (1988), who took into consideration the fluid dynamics of
relativistic shock waves. They also noted that  the resulting particle
spectral indices depend on the perturbations spectrum near the shock, in
contrast to the non-relativistic case. Kirk \& Heavens (1989) considered
the acceleration process in shocks with magnetic fields oblique to the
shock normal (see also Ballard \& Heavens 1991). They showed, contrary
to the non-relativistic results again, that such shocks led to flatter
spectra than the parallel ones. Their work relied  on the assumption of
adiabatic invariant $p_\perp^2 /B$ conservation for particles
interacting with the shock, which restricted the considerations to the
case of nearly uniform magnetic fields upstream and downstream of the
shock. A different approach to particle acceleration was presented by
Begelman \& Kirk (1990), who noted that in relativistic shocks most
field configurations lead to super-luminal conditions for the
acceleration process. In such conditions, particles are accelerated in a
single shock transmission by drifting parallel to the electric field
present in the shock. Begelman \& Kirk showed that there is more
efficient acceleration in relativistic conditions than that predicted by
the simple adiabatic theory. The acceleration process in the presence of
finite amplitude perturbations of the magnetic field was considered by
Ostrowski (1991; 1993) and Ballard \& Heavens (1992). Ostrowski
considered a particle acceleration process in the relativistic shocks
with oblique magnetic fields in the presence of field perturbations,
where the assumption $p_\perp^2 /B  = const$ was no longer valid. To
derive particle spectral indices he used a method of particle Monte
Carlo simulations and noted that the spectral index was not a monotonic
function of the perturbation amplitude, enabling the occurrence of
steeper spectra than those for the limits of small and large
perturbations. It was also revealed that the conditions leading to very
flat spectra involve an energetic particle density jump at the shock.
The acceleration process in the case of a perpendicular shock shows a
transition between the compressive acceleration described by Begelman \&
Kirk (1990) and, for larger perturbations, the regime allowing for
formation of the wide range power-law spectrum.  The analogous simulations
by Ballard \& Heavens (1992) for highly disordered background magnetic
fields show systematically steeper spectra in comparison to the above
results, as discussed by Ostrowski (1993). The particle spectrum
formation in the presence of non-linear coupling of accelerated
particles to the plasma flow has been commented by Ostrowski (1994a).
The review of the above work is presented by Ostrowski (1994b).

The shock waves propagating with relativistic velocities rise also
interesting questions concerning to the cosmic ray acceleration time
scale, $T_{acc}$. To date, however, there is only somewhat superficial
information available on that problem. A simple comparison to the
non-relativistic formula based on numerical simulations shows that
$T_{acc}$ relatively decreases with increasing shock velocity for
parallel (Quenby \& Lieu 1989; Ellison et al. 1990) and oblique
(Takahara \& Terasawa 1990; Newman et al. 1992; Drolias \& Quenby 1994;
Lieu et al. 1994; Quenby \& Drolias 1995; Naito \& Takahara 1995)
shocks. However, the numerical approaches used there, based on assuming
the particle isotropization at each scattering, neglect or underestimate
a rather significant factor controlling the acceleration process -- the
particle anisotropy. Ellison et al. (1990) and Naito \& Takahara
(1995) included also derivations applying the pitch angle diffusion
approach. The calculations of Ellison et al. for parallel shocks show
similar results to the ones they obtained with large amplitude
scattering. In the shock with velocity $0.98\,c$ the acceleration time
scale is redued on the factor $\sim 3$ with respect of the
non-relativistic formula. Naito \& Takahara considered shocks with
oblique magnetic fields. They confirmed reduction of the acceleration
time scale with increasing inclination of the magnetic field, derived
earlier for non-relativistic shocks (Ostrowski 1988). However their
approach neglected effects of particle cross field diffusion and assumed
the adiabatic invariant conservation at particle interactions with the
shock. These two simplifications limit their results to the cases with
small amplitude turbulence near the shock\footnote{One should note that
the spatial distributions near the shock derived by these authors (their
figures 1 and 2) do not show the particle density jump proved to exist
in oblique relativistic shocks by Ostrowski (1991). It is also implcitly
present in analytic derivations of Kirk \& Heavens (1989).}. One should
also note that comparing in some of the mentioned papers the derived
time scales to the non-relativistic expression does not have any clear
physical meaning when dealing with relativistic shocks.

In order to consider the role of particle anisotropic distributions and
different configurations of the magnetic field in shocks the present
work is based on the small angle particle momentum scattering approach
described by Ostrowski (1991). It enables  to model the effects of
cross-field diffusion, important in shocks with oblique magnetic fields.
Let us note (cf. Ostrowski 1993) that this code allows for a reasonable
description of particle transport in the presence of large amplitude
magnetic field perturbations also. In Section 2 below, we discuss the
differences in derivation of the cosmic ray acceleration time scale for
non-relativistic and relativistic shock waves. We demonstrate the
existence of noticeable correlations of the particle energy gains at
interactions with the shock and the respective times elapsed since the
previous interaction. Because of that any derivation of the acceleration
time scale cannot use the distribution of energy gains and the
distribution of times separately. We define the acceleration time scale
$T_{acc}^{(c)}$, as the one describing the rate of change of the cut-off
energy in the time dependent particle spectrum evolution. Then, in
Section 3, the performed simulations are described. We use the
simplified method involving small amplitude particle momentum scattering
and devoted to model the situations with anisotropic cosmic ray
distributions (Kirk \& Schneider 1987b, Ellison et al. 1990, Ostrowski
1991). In Section 4 the results are presented for shock waves with
parallel and oblique (both, sub- and super-luminal) magnetic field
configurations. We consider field perturbations with amplitudes ranging
from very small ones up to $\delta B \sim B$.  In parallel shocks
$T_{acc}^{(c)}$ diminishes with the growing perturbation amplitude and
the shock velocity $U_1$. However, it is approximately constant at a
given value of $U_1$ if we use the formal diffusive time scale as the
time unit. Another qualitative feature discovered in oblique shocks is
that due to the cross-field diffusion $T_{acc}^{(c)}$ can change with
$\delta B$ in a non-monotonic way. The acceleration process leading to
the power-law spectrum is possible in super-luminal shocks only in the
presence of large amplitude turbulence. Then, in contrast to the
quasi-parallel shocks, $T_{acc}^{(c)}$ increases with the increasing
$\delta B$. In some cases with the oblique magnetic field configuration
one may note a possibility to have an extremely short acceleration time
scales, comparable or even smaller than the particle gyroperiod in the
magnetic field upstream the shock. We also demonstrate the existence of
the coupling between the acceleration time scale and the particle
spectral index. A form of the involved relation much depends on the
magnetic field configuration. In order to evaluate some earlier
simulations applying the large angle scattering (`LAS') model and/or the
pitch angle diffusion (`PAD') model a short discussion of the results
obtained within these two approaches is presented. We stress the
importance of the proper evaluation of the effective magnetic field
(including the perturbed component) in simulations involving discrete
particle momentum scattering. In the final Section 5 we provide a short
summary and discuss some astrophysical consequences of our results. Some
preliminary results of this work were presented in Ostrowski \& Bednarz
(1995).

Below, the light velocity is used as the velocity unit, $c=1$. As the
considered particles are ultrarelativistic ones, $p = E$, we often put
the particle momentum for its energy. In the shock we label all upstream
(downstream) quantities with the index `1' (`2'). If not otherwise
indicated, the quantities are given in their respective plasma rest
frames. The shock {\em normal} rest frame is the one with the plasma
velocity normal to the shock, both upstream and downstream the shock
(cf. Begelman \& Kirk 1990). In the present paper the acceleration time
scales are always given in this particular frame.
 
\section{The acceleration time scales in non-relativistic versus
relativistic shock waves}

In the case of non-relativistic shock wave, with velocity $U_1 \ll 1$,
the acceleration time scale can be defined as

$$T_{acc} \equiv
{E \over {\overline{\Delta E} \over \Delta t}} \qquad , $$ 

\noindent
where $\overline{\Delta E}$ is the mean energy gain at particle
interaction with the shock and $\Delta t$ is the mean time between
successive interactions. One can use mean values here because any
substantial increase of particle momentum requires a large number of
shock-particle interactions and the successive interactions are only
very weakly correlated with each other. The respective expression for
$T_{acc}$ in parallel shocks,

$$T_{acc}^0 = {3 \over U_1-U_2}\, \left\{ {\kappa_1 \over 
U_1} + {\kappa_2 \over U_2} \right\} \qquad , \eqno(2.1)$$

\noindent
where $\kappa_i$ is the respective particle spatial diffusion
coefficient, has been discussed by Lagage \& Cesarsky (1983). Ostrowski
(1988) provided the analogous scale for shocks with oblique magnetic
fields and small amplitude magnetic field perturbations. It can be
written in the form

$$T_{acc}^\psi = {3 \over U_1-U_2} \, \left\{
{ \kappa_{n,1} \over U_1 \sqrt{\kappa_{n,1} \over \kappa_{\parallel,1}
\cos^2{\psi_1}} } +
{ \kappa_{n,2} \over U_2 \sqrt{\kappa_{n,2} \over \kappa_{\parallel,2}
\cos^2{\psi_2}} } \right\}
\qquad , \eqno(2.2)$$

\noindent
where the index $n$ denotes quantities normal to the shock, the index
$\parallel$ those parallel to the magnetic field, $\psi$ is an angle
between the magnetic field and the shock normal and $U_1/\cos{\psi_1}
\ll c$ is assumed. The terms $\sqrt{\kappa_n / (\kappa_\parallel
\cos^2\psi)}$ represent a ratio of the mean normal velocity of a
particle to such velocity in the absence of cross-field diffusion. One
may note that for negligible cross-field diffusion the expression (2.2)
coincides with (2.1) if we put $\kappa_{n,i}$ for $\kappa_i$ ($i$ = $1$,
$2$). The case of oblique shock with finite amplitude field
perturbations has not been adequately discussed yet, but we expect the
respective acceleration scale to be between the values given by the
above formulae for $T_{acc}^0$ and $T_{acc}^\psi$. The influence of the
particle escape boundary on the acceleration time scale and the particle
spectrum is discussed by Ostrowski \& Schlickeiser (1996).

If the shock velocity becomes relativistic, the particle energy change
at a single interaction with the shock can be comparable, or even larger
than the original energy. Moreover, after interaction with the shock,
the {\em upstream} particles with small initial angles between its
momenta and the mean magnetic field have a larger chance to travel far
away from the shock. On average, such particles spend longer times and
are able to change its pitch angles substantially until the next hits at
the shock. Then, larger pitch angles allow for particle reflections with
large energy gains or transmissions downstream (cf. Ostrowski 1994b,
also Ostrowski 1991, Lucek \& Bell 1994). Therefore, correlations of the
times between successive interactions, $\Delta t_{diff}$,  the energy
gains at these interactions, $\Delta E$, and, possibly, the probability
of particle escape occur. As an example, in Fig.~1 we map the number of
particle interaction with the shock in coordinates ($\Delta t_{diff}$,
$\Delta E$) (a more detailed discussion of the existing correlations
will be presented separately (Bednarz \& Ostrowski, in preparation)). A
cut of the presented surface at any particular value of $\Delta
t_{diff}$ gives the distribution of energy gains for particles who have
spent this time since the last interaction with the shock. A general
trend seen on the map for increasing $\Delta t_{diff}$ is growing the
value of $\Delta E / E$ for the distribution maximum. Because of that,
we propose a different approach to derivation of the acceleration time
scale with respect to that used for non-relativistic shocks. Usually the
acceleration time scale is applied for derivation of the highest
energies occurring in the particle spectrum, characterized by its
cut-off energy, $E_{c}$. Thus we use this energy scale to define the
acceleration time scale as:

$$T_{acc}^{(c)} \equiv {E_c \over \dot{E}_c } \qquad , \eqno(2.3)$$

\noindent
where $\dot{E}_c \equiv dE_c/dt$.  The rate of the cut-off energy
increase is a well-defined quantity and the time scale (2.3) has a clear
physical interpretation. The above definition does not require any limit
for the energy gains of the individual particles and all possible
correlations are automatically included here. From the meaning of the
definition (2.3) it follows that $T_{acc}^{(c)}$ is somewhat shorter
than the respective scale at the same energy for later times, required
for the respective part of the spectrum to become a pure power-law (cf.
Ostrowski \& Schlickeiser 1996). One should also note that in
relativistic shocks the time scale depends on the reference frame we use
for its measurement. In the present paper the acceleration time scales
are given in the respective normal shock rest frame. However, the
applied time units $r_{e,1}/c$ (see below) are defined with the use of
the upstream gyration time.

\begin{figure}[hbt]
\begin{center}
\mbox{\psboxto(8.5cm;5.5cm){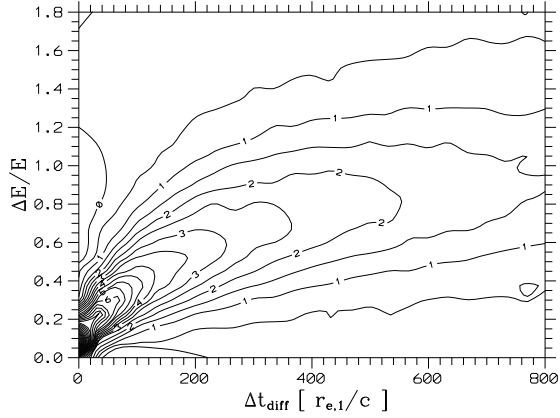}}
\caption{Distribution of particle-shock interaction events for upstream
particles, $N(\Delta t_{diff},\Delta E/E)$, in function of the upstream
diffusive times, $\Delta t_{diff}$, and the respective energy changes,
$\Delta E/E$. The correlation between $\Delta t_{diff}$ and $\Delta E/E$
is represented by a regular drift of the distribution maximum toward
higher $\Delta E/E$ when increasing $\Delta t_{diff}$. An example for
the parallel shock with $U_1 = 0.5$, $\psi_1 = 1^\circ$ and weak
scattering conditions ($\kappa_\perp / \kappa_\parallel = 1.6\cdot
10^{-6}$) is presented.}
\end{center}
\end{figure}

\section{Simulations}

We derive estimates of $T_{acc}^{(c)}$ basing on the Monte Carlo
particle simulations involving the small amplitude momentum scattering
model of Ostrowski (1991) and 2500 particles in each run of the code. In
the present considerations we discuss the role of the mean magnetic
field configuration and the amount of particle scattering. In order to
avoid the additional effects of the varying shock compression due to the
presence of different  magnetic field configurations we take the field
as a trace one, without any dynamical effects on the plasma flow. The
shock compression, as seen in the normal shock rest frame, $r$, is
derived from the approximate formulae presented by Heavens \& Drury
(1989). For illustration of the results, in the present paper we
consider the shock waves propagating in the cold electron-proton plasma.
For the {\em mean} magnetic field $B_1$ taken in the upstream plasma
rest frame and inclined at the angle $\psi_1$ with respect to the shock
normal we derive its downstream value and inclination, $B_2$ and
$\psi_2$, with the use of jump conditions presented for relativistic
shocks by e.g. Appl \& Camenzind (1988):

$$B_2 = B_1 \, \sqrt{ \cos^2{\psi_1} + R^2 \sin^2{\psi_1}}    \qquad
,  \eqno(3.1)$$

$$\tan{\psi_2} = R \, \tan{\psi_1}  \qquad , \eqno(3.2)$$

\noindent
where $R = r \, \gamma_1 / \gamma_2$ and the Lorentz factors $\gamma_i
\equiv 1/\sqrt{1-U_i^2}$ ($i$ = $1$, $2$). These formulae are valid for
both sub- and super-luminal magnetic field configurations.

We model particle trajectory perturbations by introducing small-angle
random momentum scattering along the mean-field trajectory. The particle
momentum scattering distribution is uniform within a cone wide at
$\Delta \Omega$ ($<< 1$), along the original momentum direction. The
present simulations use a constant value of $\Delta \Omega = 0.173$ ($ =
10^\circ$). Scattering events are at discrete instants, equally spaced
in time as measured in the units of the respective $r_{g,i}/c$ ($i$ =
$1$, $2$). Here we affix a gyroradius with the index `$g$' when it is a
value given for the local {\em uniform} magnetic field component. Index
`$e$' means the {\em effective} field including the field perturbations
(see below). The increasing perturbation amplitude is reproduced in
simulations by decreasing the time period $\Delta t$ between the
successive scatterings. For simplicity, we use the same scattering
pattern ($\Delta \Omega$ and $\Delta t$ in units of $r_g/c$) upstream
and downstream the shock, leading to the same values of $\kappa_\perp /
\kappa_\parallel$ in these regions (see, however, Ostrowski 1993). One
should note that the particle momentum scattering due to presence of
turbulent magnetic field is equivalent to the effective magnetic field
larger than the respective uniform mean component, $B_1$ or $B_2$. In
our model, the effective field can be estimated as

$$B_{e,i} = B_i \sqrt{1 + \left( 0.67 {\Delta \Omega \over \Delta t}
\right)^2} \quad (i = 1, 2) \qquad . \eqno(3.3)$$

\noindent
It is the lower limit for actual field since the amount of power in
perturbations with wave-lengths smaller than $c \, \Delta t$ cannot be
considered within such a simple model. Below, the derived acceleration
time scales are presented in units of the formal diffusive scale $T_0
\equiv 4( \kappa_{n,1} / U_1 + \kappa_{n,2} / U_2 ) / c$ or in
units of $r_{e,1}/c$ , in the shock normal rest frame.

Our numerical calculations involve particles with momenta systematically
increasing over several orders of magnitude. In order to avoid any
energy dependent systematic effect we consider the situation with all
spatial and time scales -- defined by the diffusion coefficient, the
mean time between scatterings and the shock velocity -- proportional to
the particle gyro-radius, $r_g = p/(eB)$, i.e. to its momentum.

For a chosen shock velocity and the magnetic field configuration we
inject particles in the shock at some initial momentum $p_0$ and follow
their phase-space trajectories. We assume the constant particle
injection to continue in time after the initial time $t_0 = 0$. As some
particles escape from the acceleration process by crossing the escape
boundary placed far behind the shock we use the trajectory splitting
procedure to keep the total amount of particles involved in simulations
constant (cf. Kirk \& Schneider 1987b; Ostrowski 1991). Here we put the
boundary at the distance $6\kappa_{2,n}/U_2 + 4 r_{g,2}$ . We checked by
simulations that any further increase of this distance does not
influence the results in any noticeable way. For every shock crossing,
the particle weight factor multiplied by the inverse of the particle
velocity normal to the shock ($\equiv$ particle density) is added to the
respective time and momentum bin of the spectrum, as measured in the
shock normal rest frame. As one considers a continuous injection in all
instants after $t_0$, in order to obtain the particle spectrum at some
time $t_j > t_0$, one has to add to particle density in a bin $p_i$ at
$t_j$ the densities in this momentum bin for all the earlier times. The
resulting particle spectra are represented as power-law functions with
the squared exponential cut-off in momentum

$$f(p,t) = A \, p^{-\alpha} \, e^{-\left( p \over p_c\right)^2} \qquad .
\eqno(3.4)$$

\noindent
In this formula three parameters are to be fitted: the normalization
constant $A$, the spectral index for the stationary solution $\alpha$,
and the momentum cut-off $p_c$ (Fig.~2; for details of the fitting
procedure see Appendix A).

\begin{figure}[hbt]
\begin{center}
\mbox{\psboxto(8.5cm;5.5cm){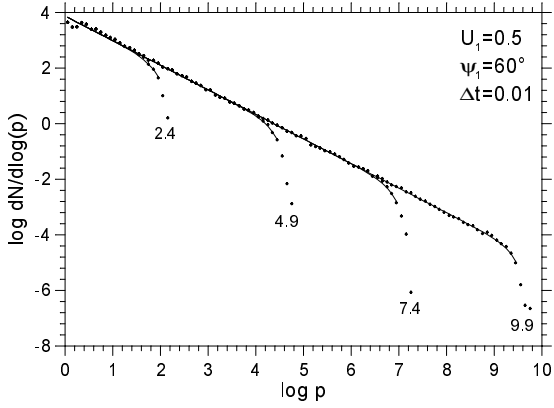}}
\caption{Particle spectra at a sequence of time instants. 
Individual simulation points represent the particle number (weight) $\Delta N$ 
per logarithmic momentum bin $\Delta \log p$. With full 
lines we present the respective fits (3.4).}
\end{center}
\end{figure}

In the simulations, due to our, proportional to momentum, scaling of the
respective quantities, the derived acceleration time scale (2.3) must be
also proportional to $p$, and thus to $r_g(p)/c$. Therefore, this time
scale measured in units of $r_g(p_c)/c$ (or $r_e(p_c) /c$) is momentum
independent and can be easily scaled to any momentum. The parameter
$T_r$ gives the value of the acceleration time scale in units of
$r_{e,1}/c$, $T_{acc}^{(c)} = T_r \, r_{e,1}/c$. The value of
$T_{acc,i}^{(c)}$ at a particular time $t_i$ is derived from the
respective values of $p_{c,i}$:

$$T_{acc,i}^{(c)} =
{p_{c,i} \over {p_{c,i}-p_{c,i-1} \over t_i-t_{i-1} }} \qquad ,
\eqno(3.5)$$

\noindent 
where we consider the advanced phase of acceleration ($p_{c,i} \gg
p_0$). As in our simulations $p_c \propto t$ the condition
$(p_{c,i}-p_{c,i-1})/p_{c,i} \ll 1$ is not required to hold in equation
(3.5). Therefore, with all scales proportional to the particle momentum,
the formula (3.5) reduces to $T_{acc,i}^{(c)} = t_i$ and the parameter
$T_r$ tends to a constant (Fig.~3). The extension of the simulated
spectra over several decades in particle energy allows to avoid problems
with the initial conditions and decrease the relative error of the
derived time scale by averaging over a larger number of instantaneous
$T_{acc,i}$.

\begin{figure}[hbt]
\begin{center}
\mbox{\psboxto(8.5cm;5.5cm){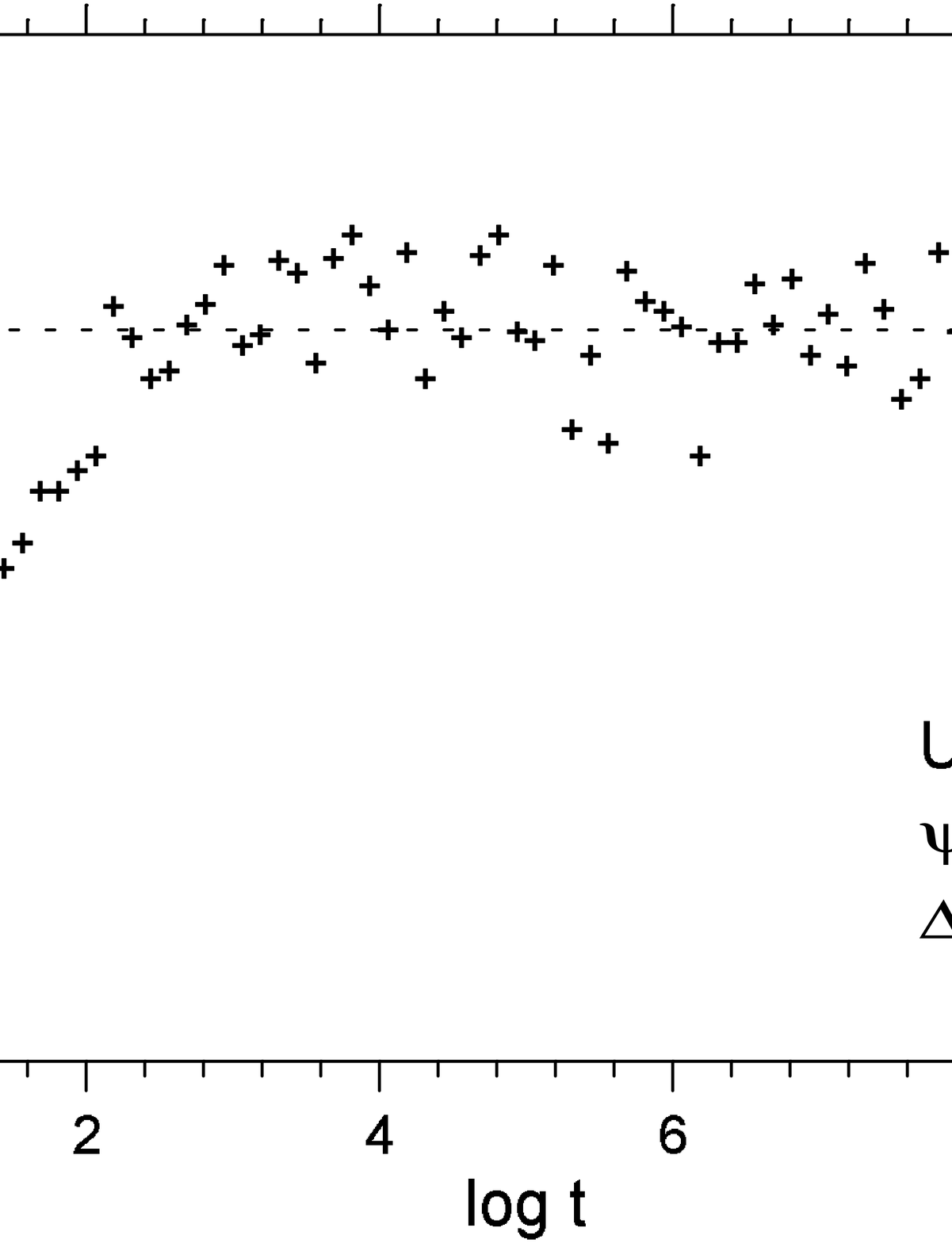}}
\caption{An example of simulated values of $T_{acc}^{(c)}$ in units of
$r_{e,1}/c$ in a single run of the code. The simulation time $t$ is
given in units of $r_{g,1}(p=p_0)/c$ in the shock normal rest frame. We
fit the final acceleration time scale only to the points at the advanced
phase of acceleration (dashed line). The dispersion of these points defines
the fitting error.}
\end{center}
\end{figure}

\section{The acceleration time scale in relativistic shock waves}

For a given relativistic shock velocity  particle anisotropy in the
shock depends on the mean magnetic field inclination to the shock normal
and the form of turbulent field. Below, we describe the results of
simulations performed in order to understand the time dependence of the
acceleration process in various conditions. In order to do that, we
consider shock waves propagating with velocities $U_1$ = $0.3$, $0.5$,
$0.7$ and $0.9$ of the velocity of light, and the magnetic field
inclinations including the quasi-parallel, oblique sub-luminal and
oblique super-luminal configurations. In all these cases we investigate
the role of varying magnitude of turbulence characterized here by the
value of $\Delta t$ or by the ratio of the diffusion coefficient across
the mean field and that along the field, $\kappa_\perp /
\kappa_\parallel$ . The relation between these parameters for $\Delta
\Omega = 10^\circ$ is presented at Fig.~4, where - at $\Delta t > 0.01$
- the presented relation has the power-law form $\kappa_\perp /
\kappa_\parallel = 6.3 \cdot 10^{-5} \, (\Delta t)^2$.

\subsection{Parallel shocks}

The most simple case for discussion of the first-order Fermi
acceleration is a shock wave with parallel configuration of the mean
magnetic field. As an example we consider the shock with negligible
field inclination $\psi_1 = 1^\circ$. For such a shock, the present
simulations do confirm the expected relation of decreasing the
acceleration time scale with increasing the shock velocity and the
amplitude of trajectory perturbations (Fig.~5). One should note at the
upper panel of the figure that for short $\Delta t$ the presented time
scales decrease more and more slowly with decreasing $\Delta t$ . It is
due to the fact that starting from some value of $\Delta t$ we reach
conditions of nearly isotropic diffusion, $\kappa_\parallel \approx
\kappa_\perp$, and further decreasing of the time delay between
scatterings decreases the acceleration time in much the same proportion
as the time unit $r_{e,1}/c$ used to measure it (cf. Equ.~3.3, Fig.~4).
In the lower panel of Fig.~5 the diffusive time scale $T_0$ is used as
the time unit. The minute differences between the successive curves
reflect the statistical fluctuations arising during simulations. Without
such fluctuations all curves should coincide. The one sigma fit errors
of $T_{acc}^{(c)}$ are indicated near the respective points. One should
note that for increasing the shock velocity the acceleration time scale
decreases with respect to the diffusive time scale.

\begin{figure}[hbt]
\begin{center}
\mbox{\psboxto(8.5cm;5.5cm){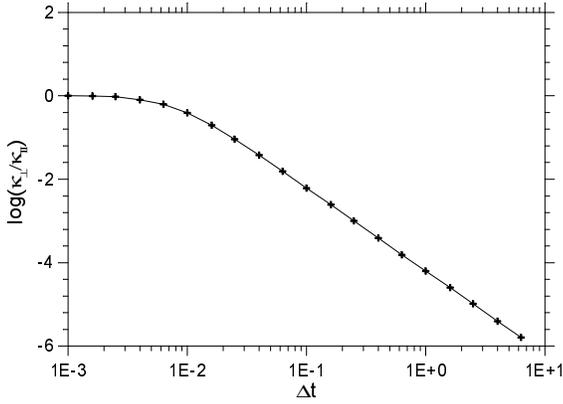}}
\caption{The relation between the scattering parameter $\Delta t$ and 
the respective ratio of diffusion coefficients $\kappa_\perp / 
\kappa_\parallel$.}
\end{center}
\end{figure}

\begin{figure}[hbt]
\begin{center}
\mbox{\psboxto(8.5cm;10cm){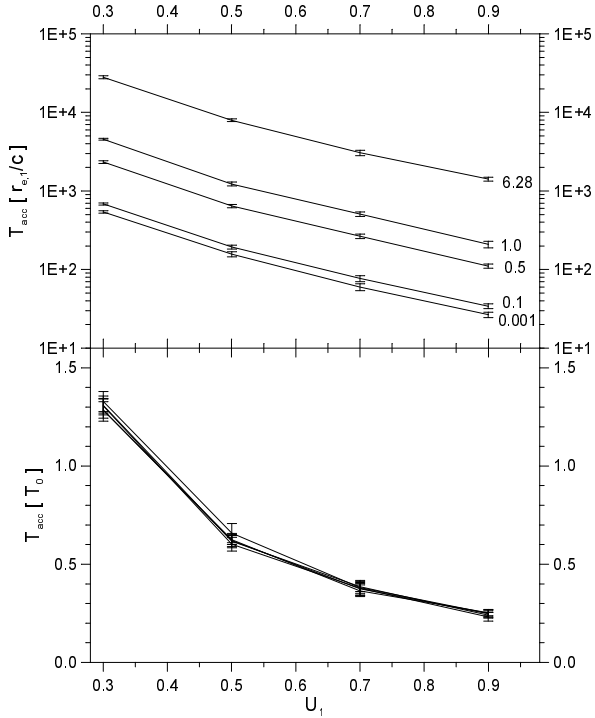}}
\caption{The values of $T_{acc}^{(c)}$ in parallel shock waves ($\psi_1
= 1^\circ$) in units of a.) $r_{e,1} / c$ and b.) $T_0$ versus the shock
velocity $U_1$. The values resulting from simulations are given for $U_1
= 0.3$, $0.5$, $0.7$ and $0.9$ with the scattering amplitude parameter
$\Delta t$ near the respective results at the upper panel.}
\end{center}
\end{figure}

\subsection{Variation of $T_{acc}^{(c)}$ with magnetic field
inclination}

In order to compare the acceleration time scales for different magnetic
field inclinations $\psi_1$ we performed simulations assuming a constant
scattering parameters upstream and downstream, yielding the same ratio
of $\kappa_\perp / \kappa_\parallel$ in these regions. However, due to
shock compression the particle gyration period is shorter downstream
than upstream, in proportion to the {\em mean} magnetic field
compression (3.1). At Fig.~6 we present the values of the acceleration
time scale derived in such conditions at different $\psi_1$. For
super-luminal shocks the results are presented for the cases allowing
for particle power-law energy spectra, i.e. when the cross-field
diffusion is sufficiently effective. Actually, the spectra with
inclinations $\alpha < 10.0$ are only included. We consider the
following values of the magnetic field inclination: $\psi_1$ = $1^\circ
$, $25.8^\circ $, $45.6^\circ $, $60^\circ $, $72.5^\circ $, $84.3^\circ
$ and $89^\circ$. The first one is for a parallel shock, the last two
ones are for perpendicular super-luminal shock with all velocities
$U_1$. The intermediate values define luminal shocks ($U_1 / \cos \psi_1
= 1.0$) at the successive velocities considered, respectively $U_1$ =
$0.9$, $0.7$, $0.5$ and $0.3$.

In general, the acceleration time scale decreases with increasing field
inclination, reaching in some cases the values comparable, or even
smaller than the particle upstream gyroperiod (6.28 in our units of
$r_{e,1}/c$). The trend can be reversed for intermediate wave amplitudes
when the magnetic field configuration changes into the luminal and
super-luminal one. Such changes are accompanied with the steepening of
the spectrum (see below). The acceleration rate at different scattering
amplitudes changes with $\psi_1$ in a way that at different inclinations
the minimum acceleration times occur at different perturbation
amplitudes (different $\Delta t$).

An important feature  of the acceleration process in relativistic shocks
should be mentioned at this point. The variations of $T_{acc}^{(c)}$ in
oblique shocks are accompanied by changes of the particle spectrum
inclination (cf. Kirk \& Heavens 1989; Ostrowski 1991). At Fig.~7, the
curves at ($T_{acc}^{(c)}$, $\alpha$) plane represent the results for
decreasing the scattering amplitude expressed with parameter $\Delta t$,
and joined with lines for the same magnetic field inclination $\psi_1$ .
For parallel shocks the changes in $T_{acc}^{(c)}$ do not lead to any
variation of the spectral index. However, for oblique sub-luminal
($\psi_1$ = $25.8^\circ$, $45.6^\circ$) and luminal ($\psi_1$ =
$60^\circ$) shocks a non-monotonic behaviour is seen. The trend in
changing $T_{acc}^{(c)}$ and $\alpha$ observed at smaller perturbation
amplitudes (larger $\Delta t$) is reversed at larger amplitudes, when
the substantial cross-field diffusion is possible.

\begin{figure*}[hbt]
\begin{center}
\mbox{\psboxto(17.8cm;12cm){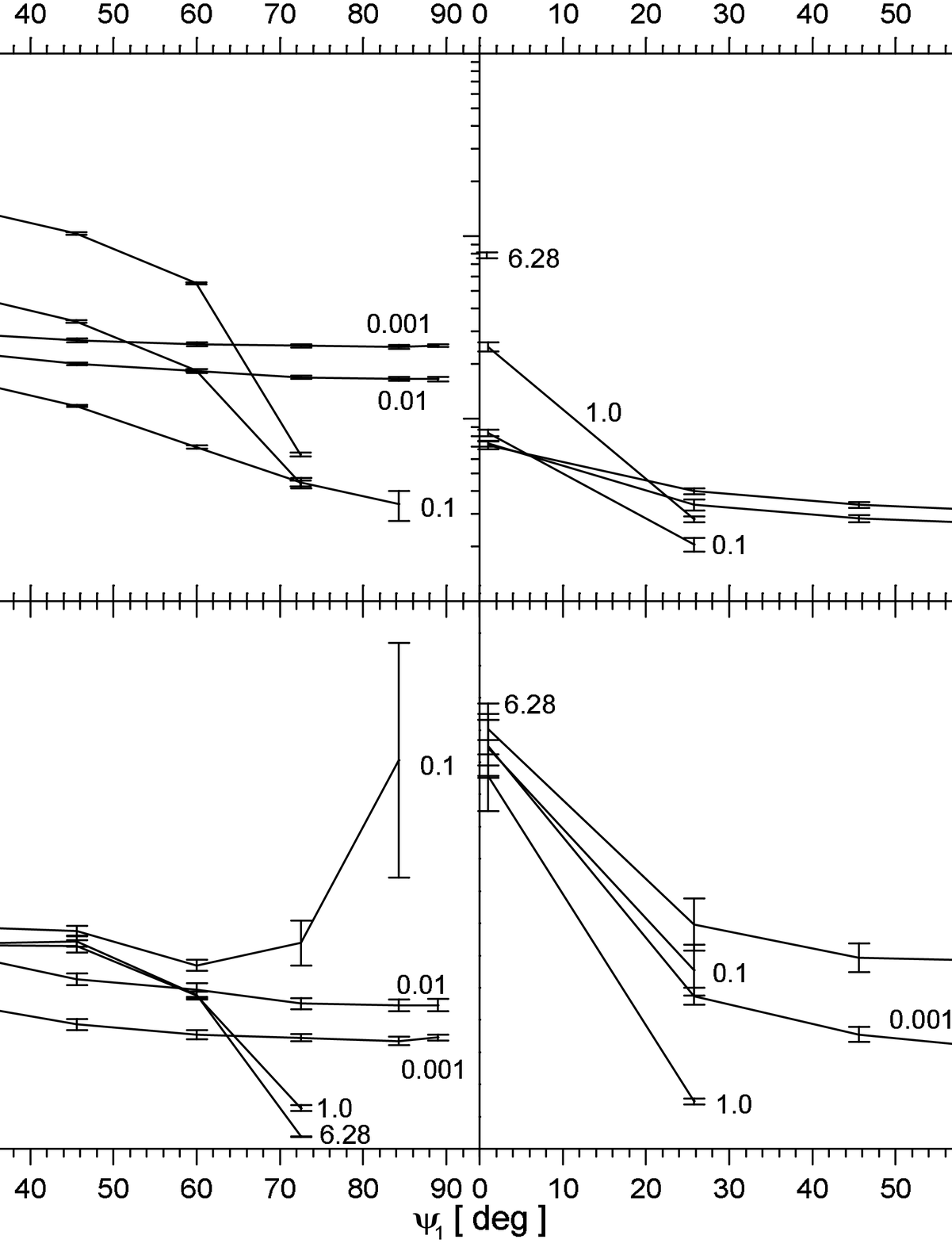}}
\caption{The value of $T_{acc}^{(c)}$ in units of $r_{e,1} / c$ (upper
panels) and $T_0$ (lower panels) versus the magnetic field inclination
$\psi_1$. The values resulting from simulations are presented for $U_1 =
0.3$ and $0.9$, with values of the parameter $\Delta t$ given near the
respective results.}
\end{center}
\end{figure*}

\subsection{Variation of $T_{acc}^{(c)}$ with varying turbulence level}

In parallel shocks the acceleration time scale reduces with the
increased turbulence level in its neighbourhood. This phenomenon, well
known for non-relativistic shocks (cf. Lagage \& Cesarsky 1983), is
confirmed here for relativistic shock velocities (Fig.~5). In general,
there are two main reasons for this change. The first one is a simple
reduction of the diffusion time of particles outside the shock due to
shorter intervals between scatterings, analogous to the decrease
observed in non-relativistic shocks. However, the increased amount of
scattering influences also the acceleration process due to changing
(decreasing) the particle anisotropy at the shock and, thus, modifying
the mean energy gain of particles interacting with the shock
discontinuity. Additionally, in oblique shocks the upstream-downstream
transmission probability may increase. One should note that the present
approach is not able to describe fully the effect of decreasing
anisotropy with the small amplitude random scattering model applied. It
is due to the fact that correlations between the successive
modifications of a trajectory (a sequence of small angle scattering acts
in this paper) in a single MHD wave cannot be accurately modelled within
the simplified approach used. A more exact approach requires integration
along the particle trajectories in realistic configurations of the
magnetic field. However, the comparison of the present simplified method
to the one involving such an integration shows a reasonably good
agreement (Ostrowski 1993) suggesting that averaging over realistic
trajectories is equivalent in some way to such averaging within our
random scattering approach.

In shocks with oblique magnetic fields a non-monotonic change of the
acceleration time scale with the amount of scattering along the particle
trajectory is observed (Fig.~8, see also Fig.~6; cf. Ostrowski 1991,
1994b for the spectral index). Increasing the amount of turbulence up to
some critical amplitude decreases the diffusion time along the magnetic
field and thus $T_{acc}^{(c)}$. However, as the mean diffusion time
outside the shock is related to the normal diffusion
coefficient\footnote{One should note that for the relativistic shocks,
due to particle anisotropy, the respective relation may be not so simple
as that given in equation (2.2) for non-relativistic shocks.} $\kappa_n$
($\kappa_{n,i} = \kappa_{\parallel,i} \cos^2{\psi_i} + \kappa_{\perp,i}
\sin^2{\psi_i}$, $i$ $=$ $1$, $2$), the increasing $\kappa_\perp$ will
lead -- for large scattering amplitudes -- to longer $T_{acc}$ in units
of [$r_e/c$] . In the units of $T_0$ the acceleration time depends only
weakly on the turbulence level and shows a small maximum for the minimum
at the presented figure. For super-luminal shocks one can note absence
of data points corresponding to low turbulence levels, where the
power-law spectrum cannot be formed or it is extremely steep. In these
excluded cases, the upstream population of energetic particles is only
compressed at the shock with the characteristic upstream time of $\sim
r_{e,1}/U_1$ (cf. Begelman \& Kirk 1990; Ostrowski 1993).

\begin{figure}[hbt]
\begin{center}
\mbox{\psboxto(8.5cm;5.5cm){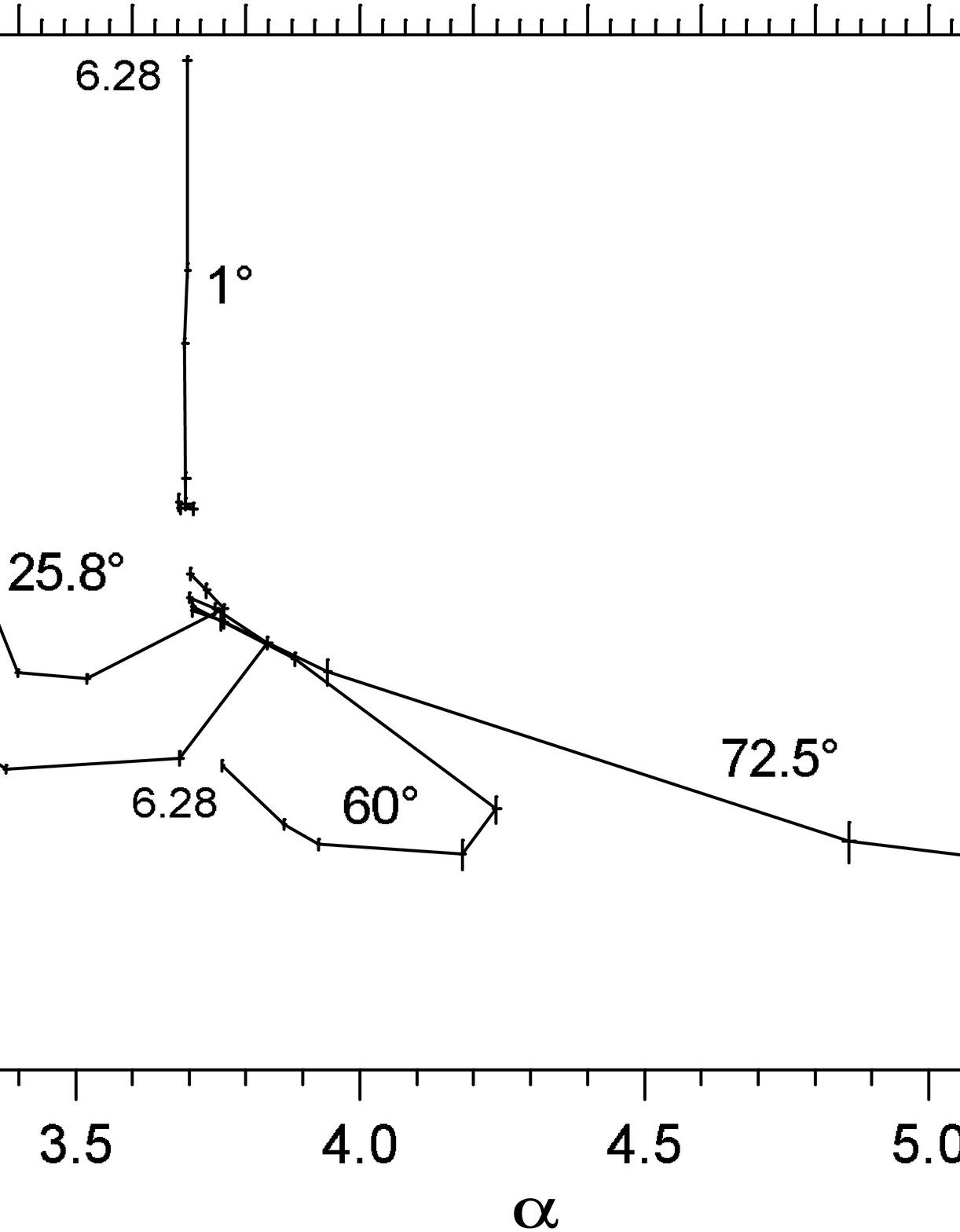}}
\caption{The values of $T_{acc}^{(c)}$ in units of $r_{e,1} / c$ at
different inclinations $\psi_1$ versus the particle spectral index
$\alpha$. The values resulting from simulations are given for $U_1$ =
$0.5$ for five values of the angle $\psi_1$ given near the respective
results. The {\em maximum} value of $\Delta t$ is given at the end of
each curve and it monotonously decreases along the curve. }
\end{center}
\end{figure}

\begin{figure}[hbt]
\begin{center}
\mbox{\psboxto(8.5cm;5.5cm){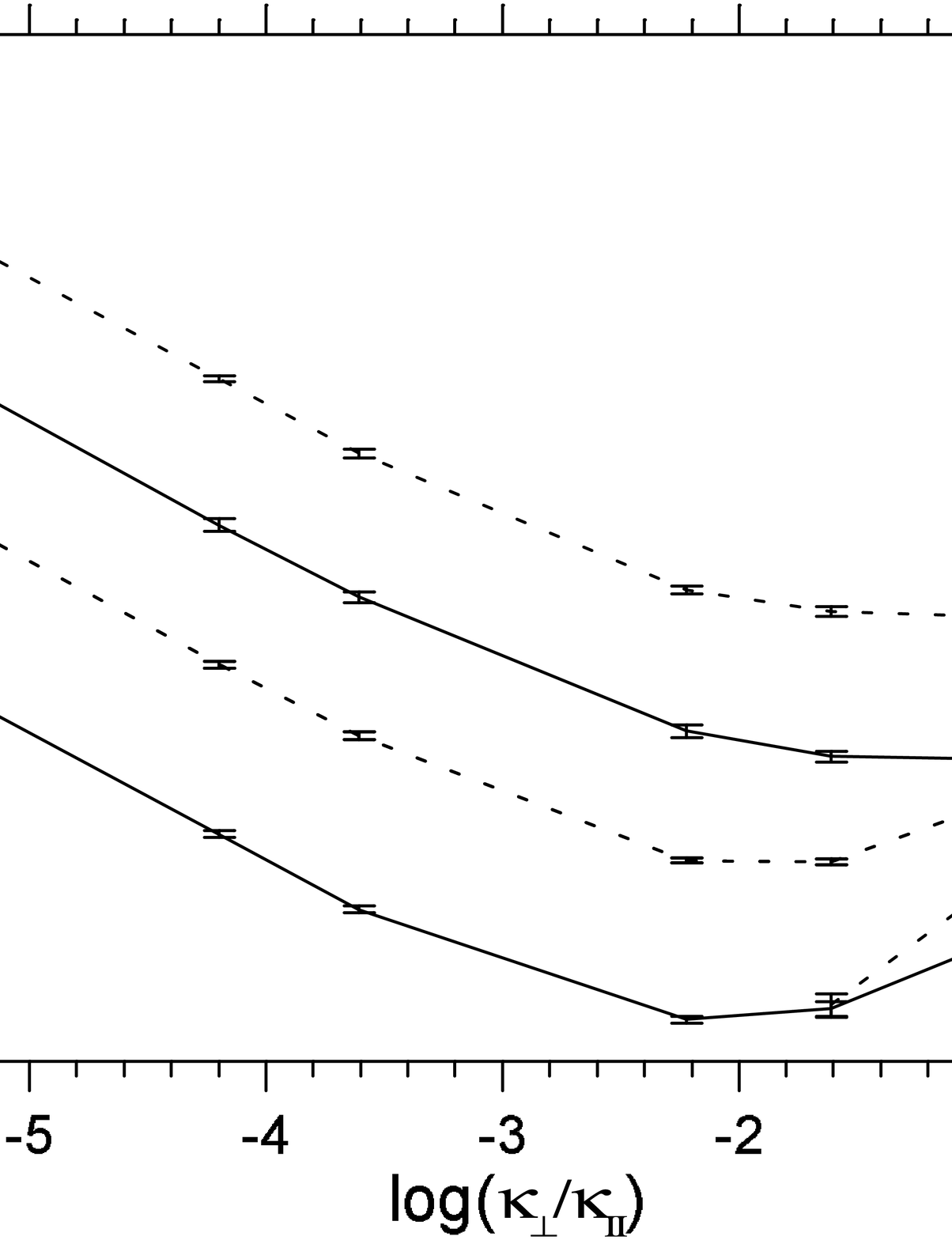}}
\caption{The acceleration time $T_{acc}^{(c)}$ versus the level of
particle scattering $\kappa_\perp / \kappa_\parallel$ for shocks with
velocity $U_1$ $ =$ $0.3$ (dashed lines) and $0.5$ (full lines). We
present results for three values of the magnetic field inclination: a.)
parallel shock ($\psi_1 = 1^\circ$), b.) a sub-luminal shock with
$\psi_1 = 45.6^\circ $ and c.) a super-luminal shock with $\psi_1 =
89^\circ $.}
\end{center}
\end{figure}

\subsection{Large angle scattering (LAS) versus pitch angle diffusion
(PAD)}

Some earlier derivations of the acceleration time scale were based on
the numerical simulations involving particle scattering at point like
scattering centres isotropizing the particle momentum at each
scattering. This approach does not provide a proper description for the
acceleration processes in shock waves moving with velocities comparable
to the particle velocity because it removes particle anisotropy and
changes the factors related to it. Moreover, against arguments presented
in some papers, such scattering pattern can not be realized in
turbulent magnetic fields near relativistic shocks, where most particles
active in the acceleration process are able do diffuse only a short distance,
below a few particle gyroradii off the shock\footnote{ However,
for the non-relativistic shock velocity and particles much above the
injection energy such approximation can be safely used (cf. Jones \&
Ellison 1991).}. Such distances are often insufficient to allow for
big particle pitch-angle changes occuring with the {\em point-like} 
scattering centres which isotropize particle momentum at each
 scattering. In shocks with oblique magnetic fields
such scattering pattern can substantially change the shape of the
accelerated particle spectrum with respect to the PAD model. 
Additionally, as an individual particle
interaction with the shock can involve a few revolutions along the
magnetic field, the usually assumed adiabatic invariant conservation,
$p_\perp^2/B = const$, cannot be valid for short inter-scattering
intervals.

Before proceeding with the results let us remark a further problem
arising within any model involving discrete particle scattering acts. As
discussed in Section 3 the presence of scattering is equivalent to the
presence of magnetic field perturbations. As a result the effective
field is larger than the uniform background field. The presented
estimate of the this field (Equ.~3.3) is valid for small angle
momentum scattering. The amount of energy in magnetic turbulence with
the waves shorter than $c\, \Delta t$ is required to be small because
the presented estimate assumes the particle momentum perturbation in
$\Delta t$ occurrs on the uniform effective perturbing field. To compare
the scattering processes with different $\Delta t$ one have to neglect
the unknown factor of the ratio of the averaged actual magnetic field to
the estimated value (like the one in Equ.~3.3). Let us note that this
factor, as well as the notion of the effective field were not considered
in the earlier papers. In derivation of the results presented below we
considered either the Equ.~(3.3) to be formally valid for both, PAD and
LAS scattering models, or we presented data with the use of units
defined by the particle gyroradius in the uniform magnetic field
component $B_0$. The later case is provided to be compared with the
earlier results by other authors.

In order to have the
comparison of the LAS versus PAD models 
meaningful we simulate the acceleration process for LAS with the
 described in Section 3 procedure, but applying the isotropic
scattering of particle momentum. Instead of our $\Delta \Omega =
10^\circ$ for the pitch angle diffusion we use  $\Delta \Omega =
180^\circ$ for the large angle scattering. The times $\Delta t$ between
successive scatterings are chosen in a way to have the same spatial
diffusion coefficients along the magnetic field for both models, as
measured in the units  related to the {\em effective}
field component ($r_e$ and $r_e/c$). With the formulae derived in simulations
($\kappa_\parallel^{PAD} = 41.7 \Delta t_{10^\circ}$ for PAD and
$\kappa_\parallel^{LAS} = 0.174 \Delta t_{180^\circ}$ for LAS) given in
the units based on $B_0$ ($r_g$ and $r_g/c$), and with the equation (3.3) for 
the effective field $B_e$ one obtaines the required
relation of the scattering times: $\Delta t_{180^\circ} = [57611 \Delta
t^2_{10^\circ} + 775.4]^{1/2}$. Let us note that as our procedure
derives the actual particle trajectories (not only the gyration centre
motion) any random scattering process results in the cross-field
diffusion and breaking the strict $p_\perp^2/B$ conservation.

\begin{figure}[hbt]
\begin{center}
\mbox{\psboxto(8.5cm;5.5cm){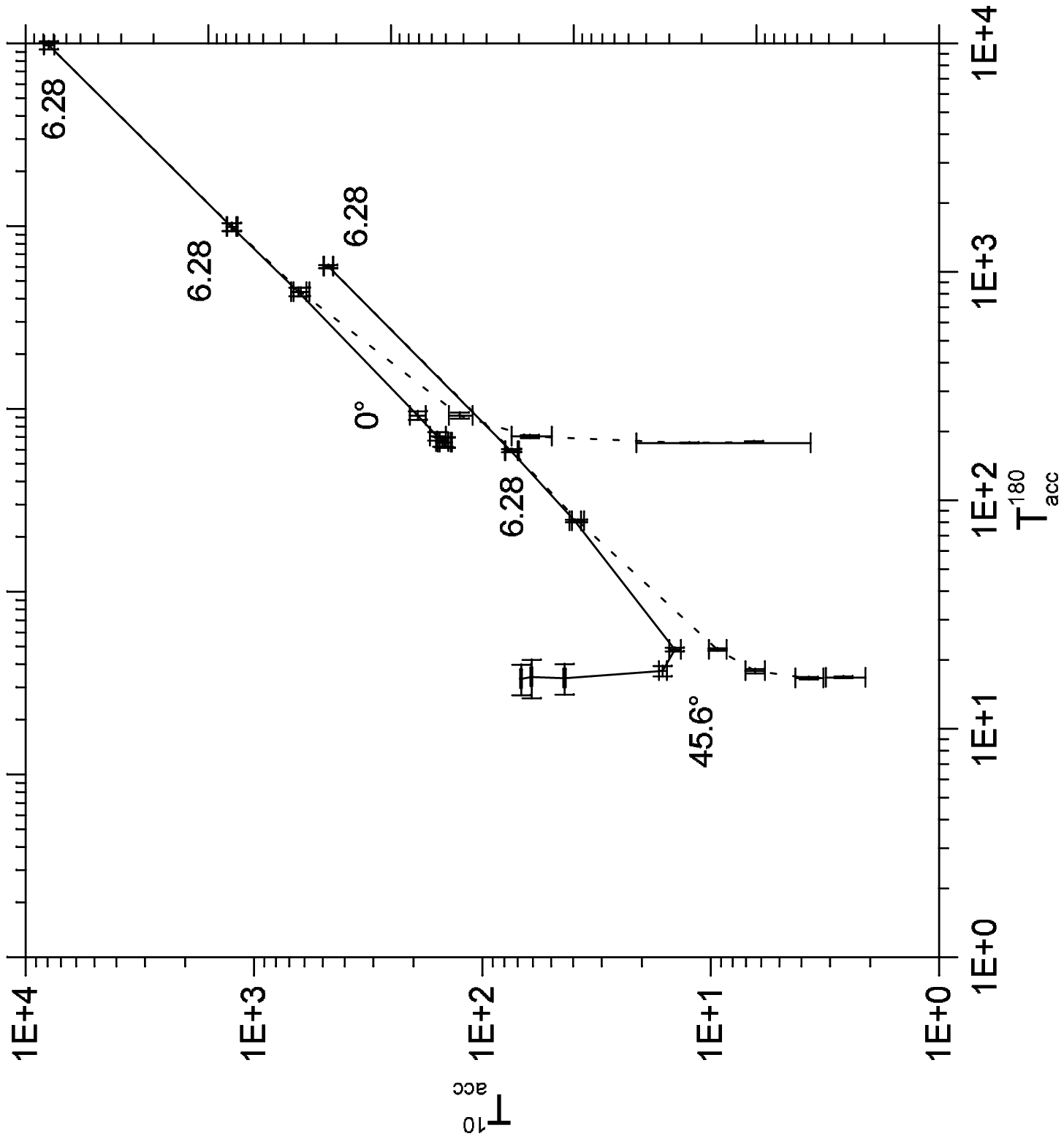}}
\caption{Comparison of the acceleration times derived with our pitch
angle diffusion procedure, $T_{acc}^{10} \equiv 
T_{acc}^{(c)}(\Delta \Omega = 10^\circ)$,
and the large angle scattering model, $T_{acc}^{180} \equiv 
T_{acc}^{(c)}(\Delta \Omega =
180^\circ)$. At the figure the points for the shock velocity $U_1 = 0.5$ 
and two magnetic field inclinations, $\Psi_1 = 0^\circ$ and $45.6^\circ$, 
are joined with lines for successive
values of $\Delta t$ . Full (dashed) lines are for the data in the units
defined by $B_e$ ($B_0$), respectively. The points for maximum $\Delta t$
($= 6.28$ in PAD) are indicated near the respective curves.}
\end{center}
\end{figure}

\begin{figure}[hbt]
\begin{center}
\mbox{\psboxto(8.5cm;5.5cm){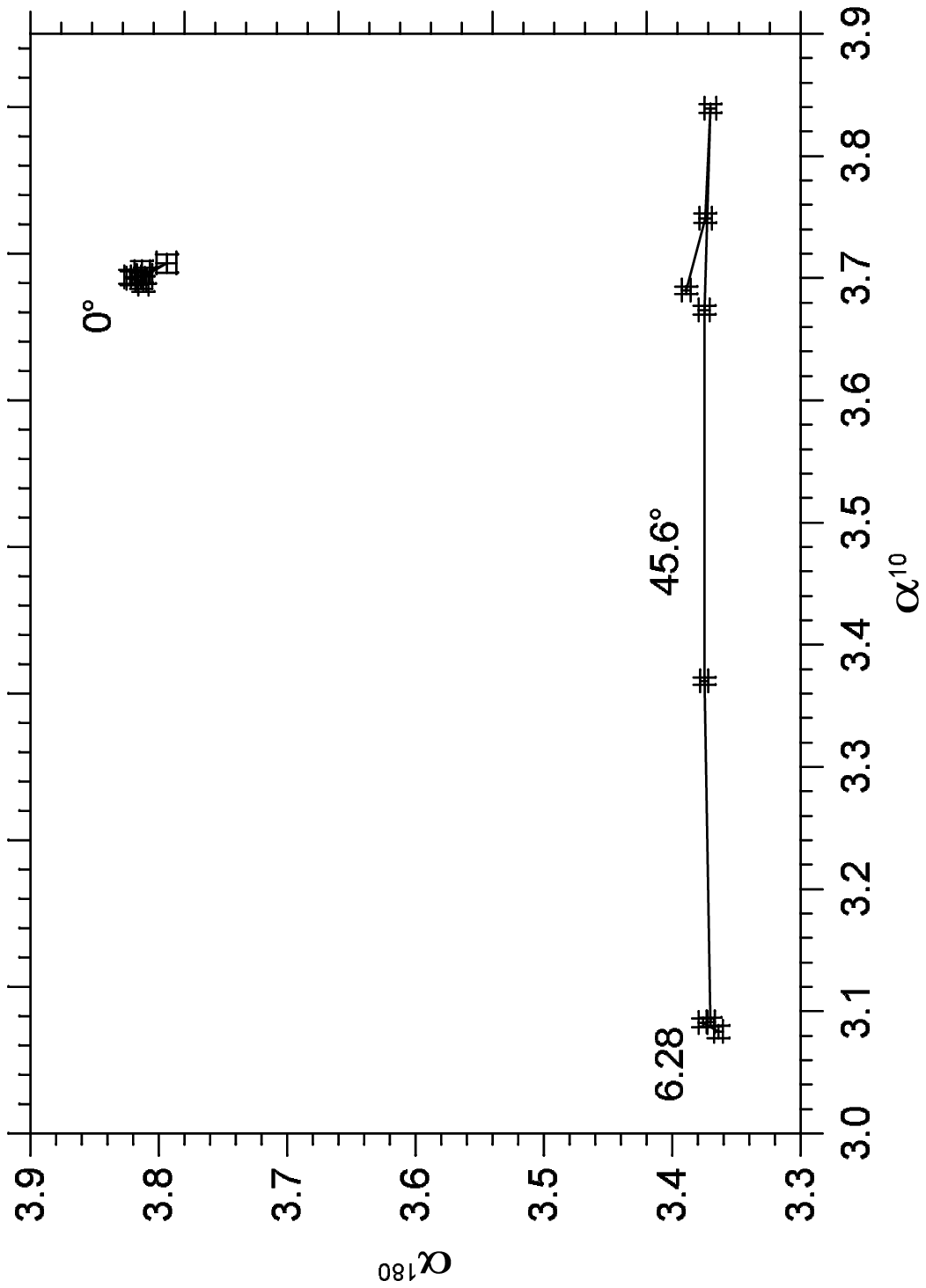}}
\caption{Comparison of the spectral indices derived with our pitch angle
diffusion procedure, $\alpha^{10} \equiv \alpha (\Delta \Omega = 10^\circ)$, 
and the large
angle scattering model, $\alpha^{180} \equiv 
\alpha (\Delta \Omega = 180^\circ)$. The data
are presented for the shock wave with $U_1 = 0.5$ and the magnetic field
inclinations $0^\circ$ and $45.6^\circ$. At the figure the data are
plotted in the same way as at Fig.~9. }
\end{center}
\end{figure}

\begin{figure}[hbt]
\begin{center}
\mbox{\psboxto(8.5cm;5.5cm){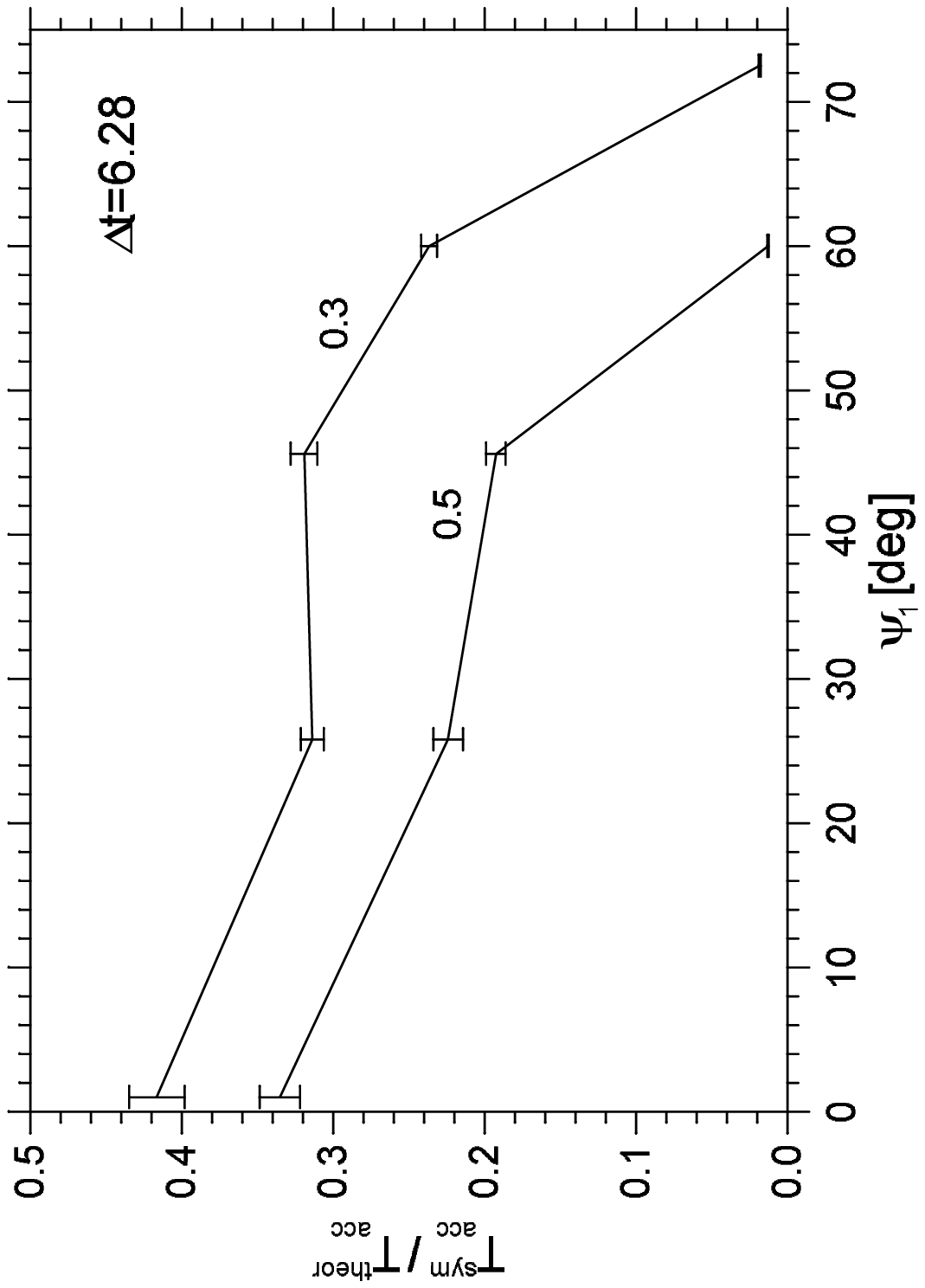}}
\caption{The ratio of the PAD simulated time scale $T_{acc}^{sym}$ to the `theoretical' 
non-relativistic time scale
$T_{acc}^{theor}$ given in Equ.~2.2 for different inclinations of the magnetic 
field $\Psi_1$. The results for $U_1 = 0.3$ and $0.5$ are presented for the case of 
negligible cross-field diffusion ($\Delta t = 6.28$ in the PAD model).}
\end{center}
\end{figure}

The discussion in this sub-section is to clarify the relation of our
present simulations to some earlier ones providing sufficient details on
the applied numerical approach. We performed simulations with the use of
the LAS model for $U_1 = 0.3$, $0.5$, $0.7$ and $0.9$, and two
inclination angles, $\psi_1 = 0.1^\circ$ and $45.6^\circ$. The results
for $U_1 = 0.5$ are presented at Fig-s~9 and 10. The comparison to the
earlier results done for the largest $\Delta t$ values, where the
effective field approximately equals the background field, shows a
general qualitative agreement with the results of Ellison et al. (1990)
and Naito \& Takahara (1995). For parallel shocks the acceleration time
scale for LAS is only slightly longer than the PAD one for all
considered shock velocities. The ratio of this time to the time scale
estimate based on the equation (2.1) for non-relativistic shocks yields
values decreasing from approximately $0.6$ at $U_1 = 0.3$ up to $0.17$
at $U_1 = 0.9$. At high velocities the ratio is much smaller (a factor 
of 2 -- 3) than the
ones derived by Ellison et al. (1990). However, there exist a
substantial difference of numerical procedures applied here and in the
mentioned paper, and we are not able to point out any single reason for
the noted discrepancy. As mentioned previously our time scale defined
with the use of the cut-off energy is shorter than the steady-state
estimate derived by Ellison et al. Additionally, in the mentioned paper
the use of analytic solutions of the spatial {\em diffusion} equation
for the case of strongly relativistic flow can be questioned. For
oblique shocks we observe analogous reduction of the acceleration time
scale as that reported by Naito \& Takahara (1995), with the PAD model
allowing for a little more rapid acceleration than the LAS model. Of
course this agreement is broken for short $\Delta t$, where the cross
field diffusion can not be neglected and the particle magnetic momentum
is not conserved at interactions with the shock. At Fig.~10 one can
compare the spectral indices derived with the use of both approaches.
As a final remark we would like to point out that the only independent
check for any numerical procedure is provided by comparying of the
derived spectral indices in the limiting cases of $\delta B << B$ and
$\kappa_\perp \approx \kappa_\parallel$ to the analytic results (e.g.
Heavens \& Drury (1988) for parallel shocks and Kirk \& Heavens (1989)
for oblique shocks). This comparison for our PAD method is presented in
Ostrowski (1991).

\section{Summary and final remarks}

In the present paper we consider the acceleration time scale in
relativistic shock waves. We argue that the numerical approach based on the
model involving particle isotropic scattering neglects or underestimates
a very significant factor controlling the acceleration process -- the
particle anisotropy. We also note that such a scattering pattern cannot
be physically realized when particle trajectory perturbations are due to
MHD waves. It is why the derivations of the present work are based on
the small-angle momentum scattering approach. We note that the code
allows for a satisfying description of particle transport in the
presence of large amplitude magnetic field perturbations as well. We
demonstrate the existence of correlation between particle energy gains
and its diffusive times. The analogous correlation is expected for the
probability of particle escape downstream the shock. Therefore, for
defining the acceleration time scale we use the rate of change of the
spectrum cut-off momentum which acommodate all such correlations. 
We performed Monte Carlo simulations for
shock waves with parallel and oblique (either, sub-luminal and
super-luminal) magnetic field configurations with different amounts of
scattering along particle trajectories. Field perturbations with
amplitudes ranging from very small ones up to $\delta B \sim B$ are
considered. In parallel shocks $T_{acc}^{(c)}$ diminishes with the
growing perturbation amplitude and the shock velocity. However, it
is approximately constant for increasing turbulence level if we use the
respective diffusive time scale as the time unit. Another feature
discovered in oblique shocks is that due to the cross-field diffusion
$T_{acc}^{(c)}$  can change with $\delta B$ in a non-monotonic way. The
acceleration process leading to the power-law particle spectrum in a
super-luminal shock is possible only in the presence of large amplitude
turbulence. Then, the shorter acceleration times occur when the
perturbations' amplitudes are smaller and the respective spectra
steeper. We discussed the coupling between the acceleration time scale
and the particle spectral index in oblique shock waves with various
field inclinations and revealed a possibility for non-monotonic
relations of these quantities.

The shortest acceleration time scales seen in the simulations are below
the particle gyroperiod upstream the shock. These times do not require
the ultra-relativistic shock velocities, but may occur in mildly
relativistic ones with the quasi-perpendicular magnetic field
configuration. One should note that - due to the larger magnetic field
downstream the shock - in this short time the particle trajectory can
follow a few revolutions near the shock with only a short section of
each one  penetrating the upstream region.

The presented estimates of the acceleration time scale provide an
interesting possibility for modelling shock waves in the conditions where
the electron spectrum cut-off energy is determined by the balance of
gains and losses. If one is able to derive the respective acceleration rate 
from the knowledge of the energy loss process, and the particle
spectral index is also known, then both these values provide constraints
for the acceleration process which could be further used to reduce the 
parameter space available for the considered shock wave (cf. Fig.~7). 

\thanks{The work of MO was partly done in Max-Planck-Institute f\"ur
Radioastronomie in Bonn. He is grateful to Prof. R. Wielebinski for a
kind invitation. The presented computations were done on the CONVEX
Exemplar SPP1000/XA-16 in ACK `CYFRONET' in Krak{\'o}w. The work was
supported by the grant PB 1117/P3/94/06 from the {\em Komitet Bada\'n
Naukowych}.}

\section*{Appendix A. Fitting the parameters $A$, $\alpha$ and $p_c$}

Any simulated spectrum was evolving in time by increasing the width of
its power-law section and, thus, the best fit of this power-law was
possible with the use of the final spectrum at maximum time. Therefore,
in the simulations we used the last spectrum to fit parameters of the
power-law normalization and the spectral index, $A$ and $\alpha$. Next,
for any earlier spectrum, these parameters were assumed to be constant
and we were fitting only the cut-off momentum, $p_c$. For each fit we
used 20 last points of the spectrum preceding the point where particle
density fell below $0.16$ of $A\,p^{-\alpha}$. The number of $0.16$ was
chosen experimentally as to obtain the best fits to the cut-off region
of the spectrum. As the distribution (3.4) represents only an
approximation to the actual particle distribution, there was no reason
to use points corresponding to lower densities, of lesser statistical
significance.

\end{document}